\documentclass[twocolumn, superscriptaddress, secnumarabic, amssymb, nobibnotes, aps, bm, pre, floatfix, longbibliography]{revtex4-2}
\usepackage[english]{babel}
\usepackage[latin9]{inputenc}
\usepackage{amsmath}
\usepackage{graphicx}
\usepackage{multirow}
\usepackage[usenames,dvipsnames]{xcolor}
\usepackage{bm}
\usepackage{mathtools}
\usepackage{amsmath}
\usepackage{rotating}
\definecolor{oceanboatblue}{rgb}{0.0, 0.47, 0.75}
\definecolor{orange}{rgb}{1,0.5,0}
\definecolor{goodgreen}{rgb}{0.1,0.5,0}
\definecolor{goodred}{rgb}{0.7,0,0}
\usepackage[colorlinks,urlcolor=goodgreen,citecolor=blue,linkcolor=goodred]{hyperref}
\usepackage{cleveref}
\usepackage{tikz}
\usepackage{tocvsec2}
\usepackage{scrextend}

\makeatletter 
\newcommand{\ii}{\text{i}}
\newcommand{\ee}{\text{e}}

\usepackage[normalem]{ulem}
\usepackage{soul}

\renewcommand{\thesubsection}{\Alph{subsection}}
\renewcommand{\thesubsubsection}{\arabic{subsubsection}}
\usepackage{braket}

\usepackage[normalem]{ulem}

\makeatother
\begin{document}
\title{Topological Properties of a Non-Hermitian Quasi-1D \linebreak Chain with a Flat Band}

\author{C. Mart\'{\i}nez-Strasser}
\email{cmartinez089@ikasle.ehu.eus}
\affiliation{Department of Physics, University of the Basque Country UPV/EHU, Apartado 644 48080, Bilbao, Spain}
\affiliation{Donostia International Physics Center (DIPC), 20018, Donostia--San Sebasti\'an, Spain}
\affiliation{Max Planck Institute for the Science of Light, Staudtstra\ss e 2, 91058, Erlangen, Germany}

\author{M. A. J. Herrera}
\email{ma.jimenez.herrera@gmail.com}
\affiliation{Department of Physics, University of the Basque Country UPV/EHU, Apartado 644 48080, Bilbao, Spain}
\affiliation{Donostia International Physics Center (DIPC), 20018, Donostia--San Sebasti\'an, Spain}

\author{A. Garc\'{\i}a-Etxarri}
\email{aitzolgarcia@dipc.org}
\affiliation{Donostia International Physics Center (DIPC), 20018, Donostia--San Sebasti\'an, Spain}
\affiliation{IKERBASQUE, Basque Foundation for Science, Plaza Euskadi 5
48009, Bilbao, Spain}

\author{G. Palumbo}
\email{giandomenico.palumbo@gmail.com}
\affiliation{School of Theoretical Physics, Dublin Institute for Advanced Studies, 10 Burlington Road, Dublin D04 C932, Ireland}

\author{F. K. Kunst}
\email{flore.kunst@mpl.mpg.de}
\affiliation{Max Planck Institute for the Science of Light, Staudtstra\ss e 2, 91058, Erlangen, Germany}

\author{D. Bercioux}
\email{dario.bercioux@dipc.org}
\affiliation{Donostia International Physics Center (DIPC), 20018, Donostia--San Sebasti\'an, Spain}
\affiliation{IKERBASQUE, Basque Foundation for Science, Plaza Euskadi 5
48009, Bilbao, Spain}

\begin{abstract}
The spectral properties of a non-Hermitian quasi-1D lattice in two of the possible dimerization configurations are investigated. Specifically, it focuses on a non-Hermitian diamond chain that presents a zero-energy flat band. The flat band originates from wave interference and results in eigenstates with a finite contribution only on two sites of the unit cell. To achieve the non-Hermitian characteristics, the system under study presents non-reciprocal hopping terms in the chain. This leads to the accumulation of eigenstates on the boundary of the system, known as the non-Hermitian skin effect. Despite this accumulation of eigenstates, for one of the two considered configurations, it is possible to characterize the presence of non-trivial edge states at zero energy by a real-space topological invariant known as the biorthogonal polarization. This work shows that this invariant, evaluated using the destructive interference method, characterizes the non-trivial phase of the non-Hermitian diamond chain. For the second non-Hermitian configuration, there is a finite quantum metric associated with the flat band. Additionally, the system presents the skin effect despite the system having a purely real or imaginary spectrum. The two non-Hermitian diamond chains can be mapped into two models of the Su-Schrieffer-Heeger chains, either non-Hermitian, and Hermitian, both in the presence of a flat band. This mapping allows to draw valuable insights into the behavior and properties of these systems.
\end{abstract}
\date{\today}
\maketitle

\renewcommand{\thesection}{\arabic{section}} %
\renewcommand{\thesubsection}{\arabic{section}.\arabic{subsection}} %
\renewcommand{\thesubsubsection}{\arabic{section}.\arabic{subsection}.\arabic{subsubsection}} %

\section{Introduction}  \label{sec:intro}

Non-Hermitian (NH) physics is an emergent field of research that has important implications both for quantum and classical physics~\cite{Bergholtz_2021,Ashida2020,Ding_2021,Okuma_2022,Wang_2023,Ghatak_2020}. A systematic study of this field started with the cornerstone works by Bender and Boettcher on Hamiltonian systems preserving the combination of parity and time-reversal ($\mathcal{PT}$) symmetry, ensuring a real spectrum~\cite{Bender_1998,Bender_2007}. At the moment, model Hamiltonian systems respecting $\mathcal{PT}$ symmetry are considered  excellent models for describing dissipative systems with balanced gain and loss in an effective way~\cite{Bergholtz_2021}.

The condition of reality of the spectrum can be extended by considering a more general symmetry class known as \emph{pseudo-Hermiticity} that includes the $\mathcal{PT}$-one~\cite{mostafazadeh2002pseudo1}.
In general, NH operators exhibit intriguing phenomena such as non-orthogonal eigenstates and complex energy spectra containing exceptional points (EPs), representing stable points of band degeneracies at which not only the eigenvalues but also the eigenvectors coalesce~\cite{Bergholtz_2021,Ding_2021,Okuma_2022}.

Recent research has focused on the topological characterization of NH systems~\cite{Bergholtz_2021,Ding_2021,Okuma_2022}, expanding upon the framework established for Hermitian condensed-matter systems. One of the key effects of moving to the NH realm leads to the extension of the topological classification, considering now that  complex conjugation and transposition are no longer equivalent for non-Hermitian Hamiltonians,
moving to 38  classes~\cite{Kawabata_2019} instead of the tenfold classification of the Hermitian counterpart~\cite{Chiu_2016}.

A key difference between NH and Hermitian systems is the breakdown of the traditional bulk-boundary correspondence (BBC), which predicts the appearance of boundary modes based on bulk topological invariants. This breakdown can be manifested through the non-Hermitian skin effect, where bulk states accumulate at the edges of the system~\cite{yao2018edge, Zhang_2022c,Lin_2023}. Two possible research lines have been developed for reestablishing the BBC. The first one is based on the biorthogonal bulk-boundary correspondence approach~\cite{kunst2018biorthogonal}; here, right and left eigenvectors of the system under open boundary conditions (OBCs) are combined to project the boundary mode localization and predict gap closings accurately. The second method is based on the concept of the so-called \emph{generalized Brillouin zone} (BZ)~\cite{yao2018edge, Yao2018, yokomizo2019non}, in which additional information is encoded inside the standard Bloch bands.

For the investigation of topological effects in NH systems, 1D systems are ideal platforms for presenting the key features~\cite{Hatano_1996,lee2016anomalous,Lieu_2018,kunst2018biorthogonal,edvardsson2020phase}. These simple 1D models exhibit many of the unusual properties of NH systems. A paradigmatic example is the Hatano-Nelson model~\cite{Hatano_1996}, which is a one-band system with anisotropic nearest-neighbor couplings originally proposed to study localization transitions in superconductors. Under periodic boundary conditions (PBCs), this model features loops in the complex spectrum resulting in a non-trivial spectral winding number~\cite{Gong2018, Shen2018}. When going to OBCs, this translates into the appearance of the NH skin effect, thus establishing a new, truly NH bulk-boundary correspondence ~\cite{Borgnia_2020,Okuma2020,Zhang2020}.

Even richer NH phenomena can be observed in NH versions of the Su-Schrieffer-Heeger (SSH) chains \cite{su1979solitons, su1980soliton, heeger1988solitons}. This 1D two-band system features zero-energy end modes in the Hermitian case, which are topologically protected by a non-trivial winding number. A NH version of this chain with asymmetric hopping has been shown to host zero-energy boundary states as well as NH skin states \cite{yao2018edge, kunst2018biorthogonal}. As such, this system breaks the conventional BBC and needs to be treated either in the biorthogonal picture~\cite{kunst2018biorthogonal} or within the framework of the generalized BZ \cite{yao2018edge}. A $\mathcal{PT}$-symmetric version of the SSH chain has also been studied, which features an onsite complex potential with alternating sign ~\cite{Lieu_2018, Halder_2022}. In this case, the boundary states acquire an imaginary energy, while the NH skin effect is absent, such that the traditional BBC applies.

In this work, we present the topological properties of a quasi-1D system: the diamond chain~(DC)~\cite{Vidal_2000,Bercioux_2004,Bercioux_2005,Bercioux_2017}. The unit cell of this quasi-1D system contains three sites with unequal connectivity; in the following, we will name the site with higher connectivity as H, whereas the sites with lower connectivity will be named A and B | see Fig.~\ref{fig1}. This lattice model has also been studied with respect to the effects of localization due to an external magnetic field and many-body effects~\cite{Vidal_2000,Rizzi_2006,Cartwright_2018}. The imbalance in the connectivity results in the appearance of a bulk zero-energy mode, where the wave function is localized only in the sites of lower connectivity with opposite amplitudes, while it has zero amplitude on the remaining H sites. In the Hermitian case, in Ref.~\cite{Bercioux_2017}, it was shown that two possible dimerizations could be chosen for the DC, but only one of these presents topological properties analogous to the SSH model~\cite{su1979solitons, su1980soliton, heeger1988solitons}. Possible experimental implementations for the Hermitian system involve cold atoms in optical lattices~\cite{Hyrk_s_2013}, photonic~\cite{Mukherjee_2018}, and solid-state platform~\cite{Huda_2020}. Within this work, we investigate two possible NH configurations of the DC chain.
A NH version of the DC has already been investigated with particular emphasis on possible photonic realizations and focusing mainly on the $\mathcal{PT}$-symmetric version~\cite{Leykam_2017,Ke_2019,Zhang_2020,Ding_2021,Parkavi_2021,Parkavi_2022,Zhang_2022b}. Additional research has investigated the possibility of obtaining lasing from the flat band~\cite{Amelio_2023}. Within this work, we will relax this symmetry constriction.

We introduce non-Hermiticity by imposing a preferred hopping direction within unit cells, resulting in two non-reciprocal tight-binding models. The motivation beyond studying the NH diamond lattice stems from its potential for realizing two distinct dimerization configurations within the lattice. The first configuration displays zero-energy edge states, which can be characterized through the evaluation of the biorthogonal polarization~\cite{kunst2018biorthogonal, edvardsson2020phase}. The interest in the second configuration arises from the properties of the flat band~\cite{shu2022suppression,Bartlett_2021,xu2020non} resulting in a  \emph{giant boost} of the quantum metric properties~\cite{Bouzerar_2022}. Additionally, we show how to map these two systems into a combination of Hermitian and NH SSH models coupled to a flat band. This mapping allows us to draw valuable insights into the behavior and properties of these systems, opening new avenues for further exploration of non-Hermitian 1D systems.

This article is structured as follows: in Sec.~\ref{sec_model}, we describe the NH DC in the two possible dimerization configurations that we have uncovered.  In Sec.~\ref{sec_topo}, we explore the topological properties of these two NH lattice configurations, focusing in particular on the biorthogonal polarization and the quantum metric. In Sec.~\ref{sec_rot}, we present the path toward the unitary transformation of the lattices into SSH chains, plus an extra site describing the flat band. We conclude in Sec.~\ref{Conclusions} with a summary of our findings. We include technical appendices: App.~\ref{App1} presenting an analytical expression for the system wave functions in the case of translational invariance, and App.~\ref{App2} summarizing the symmetry properties is the two lattices at the end of our work.

\section{Results}\label{sec_results}
%
%
\subsection{Systems and Formalisms}\label{sec_model}
%
%
\begin{figure}
    \centering
    \includegraphics[width=\columnwidth]{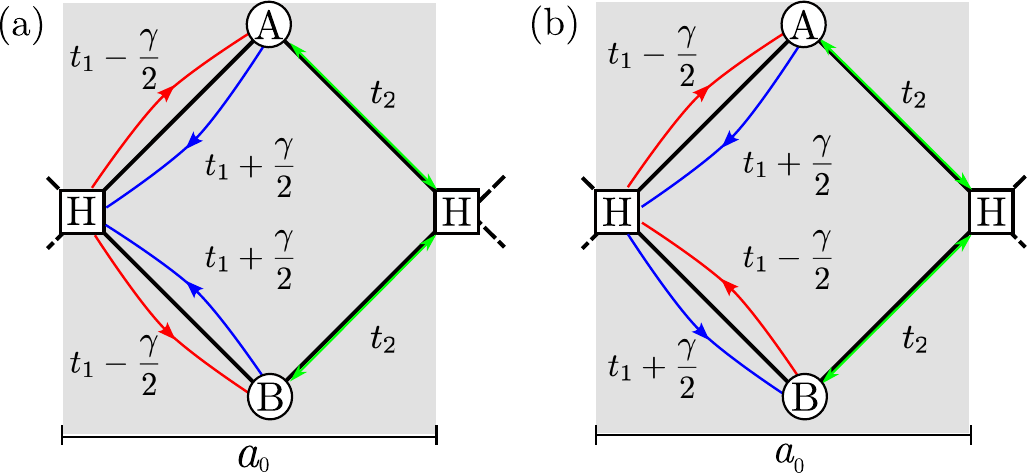}
    \caption{\label{fig1} Sketch of the non-Hermitian diamond lattice in the A and B configurations, panel (a) and (b), respectively. The gray area denotes the unit cell of each lattice configuration. The lattice periodicity is $a_0$. In both panels, the blue arrows correspond to a hopping term of strength $t_1+\frac{\gamma}{2}$, whereas the red ones to $t_1-\frac{\gamma}{2}$.}
\end{figure}
%
%
\subsubsection{Spectrum and symmetries}\label{subsec:ss}

The non-Hermitian counterpart of the diamond chain under ``lattice I" configuration from Ref.~\cite{Bercioux_2017} can be divided into two different systems according to their NH coupling configuration --- DCA and DCB. The main differences imply an intracell hopping from site H to site B of $t_1-\gamma/2$ in DCA and $t_1+\gamma/2$ in DCB. We present a sketch of these two lattices in Fig.~\ref{fig1}. In the NH version, both lattice systems still present a zero-energy flat band in the energy spectrum; this originates in the unequal connectivity between the three lattice sites in the unit cell | H connected to four neighbors, and A and B connected to two. The two chains could be thought of as two joined NH SSH models~\cite{Lieu_2018}; however, the sharing of a common lattice site (H) drastically changes the spectral properties of both the Hermitian and NH systems.
%
%
\begin{figure}[!t]
    \centering
    \includegraphics[width=\columnwidth]{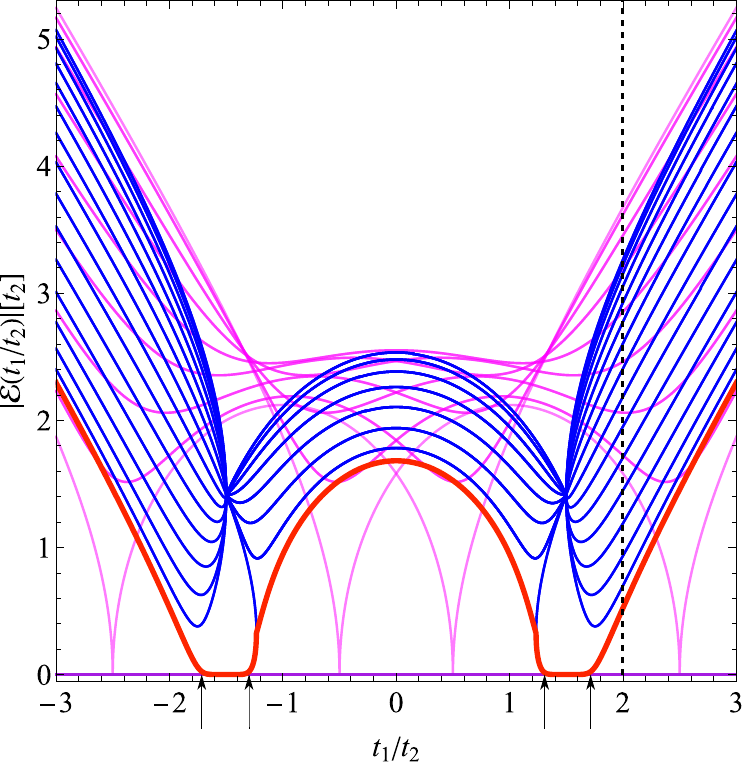}
    \caption{\label{fig2} The absolute value of the eigenenergies is examined as a function of the hopping parameters, denoted as $t_1$, in the non-Hermitian DCA model with a fixed asymmetry term $\gamma=3$ and a system size $N=16$. The pink lines represent the bulk spectra obtained under periodic boundary conditions (PBC), and the blue (red) lines correspond to the bulk (edge) spectra obtained under open boundary conditions (OBC), signaling the phase-transition points through arrows. The zero-energy flat band, depicted in purple, is present in both the PBC and OBC scenarios. Moreover, the black dashed line at $t_1/t_2=2$ indicates the hopping parameter values at which the skin effect in Fig.~\ref{fig3} has been computed.}
\end{figure}
%
%
\subsubsection{Diamond Chain A (DCA)}\label{subsubsec:L1A}
The tight-binding Hamiltonian of the non-Hermitian diamond chain in the DCA configuration [see Fig.~\ref{fig1}a)] reads: 
%
%
\begin{align}\label{1} 
\mathcal{H}_{\mathrm{DCA}}=\sum_{n} &\left\{\left[t_2(c_{\mathrm{A},n}^{\dagger}+c_{\mathrm{B},n}^{\dagger})c_{\mathrm{H},n+1}+\mathrm{h.c.}\right]\right.\nonumber\\
&\left.+\left(t_1-\frac{\gamma}{2}\right)(c_{\mathrm{A},n}^{\dagger}+c_{\mathrm{B},n}^{\dagger})c_{\mathrm{H},n} \right.\\
&\left.+\left(t_1+\frac{\gamma}{2}\right)c_{\mathrm{H},n}^{\dagger}(c_{\mathrm{A},n}+c_{\mathrm{B},n})\right\}\nonumber
\end{align}
%
%
The operators $c_{\alpha,n}^{\dagger}$ and $c_{\alpha,n}$ create and annihilate a state on sub-lattice site $\alpha \in \{\mathrm{A,B,H}\}$ of unit cell $n$, respectively. Here, $n$ ranges from $1$ to $N$, where $N$ is the total number of unit cells. The parameters $t_1$ and $t_2$ represent the intracell and intercell hopping parameters, respectively, while $\gamma$ is an asymmetry term that introduces the non-Hermitian character into the system | see Fig.~\ref{fig1}a. Throughout this paper, we fix all these parameters to be real-valued, i.e., $\{t_1,t_2,\gamma\} \in \mathbb{R}$.

%
%
\begin{figure}[!t]
    \centering
    \includegraphics[width=\columnwidth]{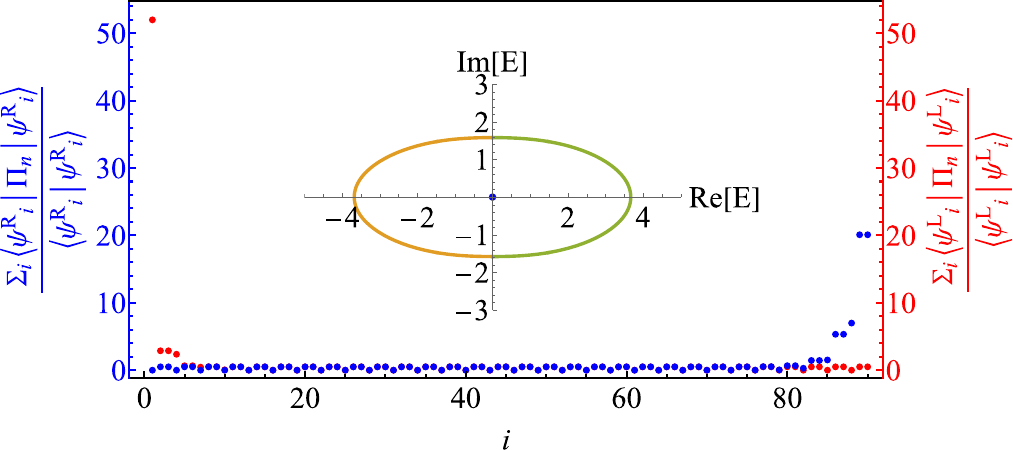}
   \caption{Skin effect of the non-Hermitian DCA chain represented through the sum of the squared amplitudes of each site $i$ with $N=30$, $\gamma=3$ and $t_1/t_2=2$ To visualize the eigenstates, we are considering the base $(\mathrm{H}_n,\mathrm{A}_n,\mathrm{B}_n)$. Additionally, we are rescaling their expectation value for comparative reasons. Inset: complex plane with the corresponding positive(green), negative(yellow) and flat(blue) energy bands. }
    \label{fig3}
\end{figure}
%
%
Assuming translational invariance, the NH Hamiltonian operator can be written in reciprocal space as
%
%
\begin{equation}\label{2}
h_\mathrm{DCA}(\kappa)=d_x\Sigma_x+d_y\Sigma_y
\end{equation}
%
%
after an overall rotation $\text{e}^{\text{i}\kappa/2}$ over H sites in reciprocal space, where $\kappa=k a_0$  is a real and dimensionless quasi momentum, $a_0$ is the lattice periodicity constant, and the momentum $k\in\textrm{BZ}$. We have defined the  matrices $\Sigma_x$ and $\Sigma_y$ as:
%
%
\begin{subequations}\label{sigmamatrices}
\begin{align}\label{3}
    \Sigma_x=\frac{1}{\sqrt{2}}
     \begin{pmatrix}
        0 & 1 & 0 \\
        1 & 0 & 1 \\
        0 & 1 & 0 \\
    \end{pmatrix}
    ~ \text{and} ~ 
    \Sigma_y=\frac{1}{\sqrt{2}}
    \begin{pmatrix}
        0 & \text{i} & 0 \\
        -\text{i} & 0 & -\text{i} \\
        0 & \text{i} & 0 \\
    \end{pmatrix}
\end{align}
%
%
accompanied by 
%
%
\begin{align}\label{4}
    \Sigma_z=\frac{1}{2}
    \begin{pmatrix}
         1 & 0 & 1 \\
         0 & -2 & 0 \\
         1 & 0 & 1 \\
    \end{pmatrix}
   ~ \text{and} ~ 
    \tilde{\mathbb{I}}=\frac{1}{2}
    \begin{pmatrix}
        1 & 0 & 1 \\
        0 & 2 & 0 \\
        1 & 0 & 1 \\
    \end{pmatrix}
\end{align} 
\end{subequations}
%
%
Therefore, the characteristic $\boldsymbol{d}$-vector, $\boldsymbol{d}=(d_x,d_y,d_z)$ reads explicitly
%
%
\begin{equation}\label{dvector}
 \boldsymbol{d}(\kappa)= \sqrt{2}(t_1 +t_2\cos(\kappa),t_2\sin(\kappa)+\text{i}\gamma/2,0)
\end{equation}
%
%
It has to be noted that the set $\{\tilde{\mathbb{I}},  \Sigma_x, \Sigma_y, \Sigma_z\}$ forms an orthogonal base with SU(2) Lie algebra with $[\Sigma_n,\Sigma_m]=2\ii \epsilon_{nmk} \Sigma_k\tilde{\mathbb{I}}$  and $\{\Sigma_n,\Sigma_m\}=2\delta_{nm}\tilde{\mathbb{I}}$, where $\epsilon_{nmk}$ is the Levi-Civita tensor with $n,m,k\in\{x,y,z\}$, and $[.,.]$ and $\{.,.\}$ are the commutator and the anticommutator, respectively~\cite{Bercioux_2017}.

The representation in Eq.~\eqref{2} results in a three-band model. This can be considered equivalent to the one obtained for the NH SSH two-band model~\cite{kunst2018biorthogonal,koch2020bulk} with the inclusion of a zero-energy flat band. This analogy will be made more explicit in Sec.~\ref{sec_rot}.

The energy spectrum of DCA is
%
%
\begin{align}
   E^{\mathrm{DCA}}_\alpha &=\alpha\sqrt{2 \left(t_1^2+t_2^2\right)+4 t_1 t_2 \cos (\kappa)+2 \text{i} \gamma  t_2 \sin (\kappa)-\frac{\gamma^2}{2}}\label{spectrumDBA}
\end{align}
%
%
with $\alpha\in\{0,\pm\}$. We note in passing that this spectrum is identical to the NH-SSH model up to a factor $\sqrt{2}$~\cite{kunst2018biorthogonal}, see Eq.~\eqref{dvector}.
We present in App.~\ref{App1} the analytical expression of the eigenstates of Eq.~\eqref{2}.

By inspecting the Jordan decomposition of the DCA Hamiltonian and the spectrum in Eq.~\eqref{spectrumDBA}, we find four exceptional points at $\mathrm{Im}[E_{\pm}]=\mathrm{Re}[E_{\pm}]=0$. These points are located at $t_1=-t_2\pm\gamma/2$ for $\kappa=0$ and $t_1=t_2\pm\gamma/2$ for $\kappa=\pi$, recognizing the similarity with the standard non-Hermitian SSH model~\cite{kunst2018biorthogonal}. As $\kappa$ ranges from 0 to 2$\pi$, the system's energy spectrum on the complex energy plane is formed of two bands and a zero-energy point that signals the zero-energy flat band of the system under OBC (see the inset of Fig.~\ref{fig3}). Depending on the choice of parameters, the two-band energies can either braid into two separate loops (inside which a reference point can be placed acting as a point gap) or into a single loop (also presenting a point gap)~\cite{Bartlett_2021}. Both phases display a qualitative change in the spectrum, in which the phase transitions correspond to the crossings of exceptional points at which the Hamiltonian becomes defective. We find a real and a fully imaginary gap for $\kappa=\{0,\pi\}$ with $|\gamma/2| \leq |t_1\pm t_2| $ and $|\gamma/2| \geq |t_1\pm t_2|$, respectively.

%
%
\begin{figure}[!t]
    \centering
    \includegraphics[width=\columnwidth]{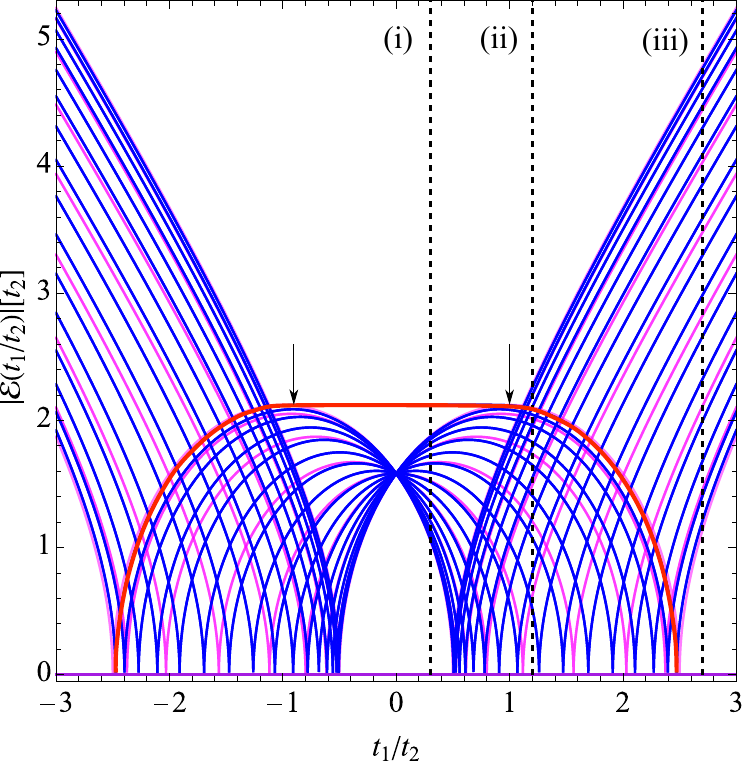}
    \caption{Absolute value of the eigenenergies as a function of the hopping parameters $t_1$ of the non-Hermitian DCB model for $\gamma=3$ and $N=16$. The pink lines depict the bulk spectra under periodic boundary conditions. The  blue (red) lines correspond to the bulk (edge) spectra obtained under open boundary conditions (OBC), signaling the phase transition points through arrows. The black dashed lines indicate the specific values of the hopping parameters i) $t_1/t_2=0.3$, ii)~ $t_1/t_2=1.2$ and iii)~ $t_1/t_2=2.7$ at which the corresponding skin effects have been computed (see Fig.~\ref{fig5}). The zero-energy flat band is depicted in purple, present under both boundary conditions.}
    \label{fig4}
\end{figure}
%
%
The finite-size system presents edge modes under the appropriate choice of system parameters. In Fig.~\ref{fig2}, we compare the energy spectrum |now denoted through $\mathcal{E}$| of a finite-size system with open boundary conditions, in which edge modes appear, and with periodic boundary conditions. In Sec.~\ref{sec_topo}, we will characterize the presence of these edge states using the biorthogonal polarization~\cite{kunst2018biorthogonal}, enabling us to differentiate them from bulk eigenstates that pile up at the boundaries; a topologically trivial effect in non-Hermitian systems lacking parity symmetry. This is known under the name of non-Hermitian \emph{skin} effect and corresponds to a piling of the system eigenstates at the boundary of the system. We show in Fig.~\ref{fig3} the skin effect for left and right eigenstates for the DCA system. Upon evaluating the Jordan canonical form of the non-Hermitian DCA system at $t_1=\gamma/2$, we find  $N+3$ independent eigenvectors, $N$ corresponding to the zero-energy flat band and three corresponding to EPs. Similar to the non-Hermitian SSH~\cite{alvarez2018non}. From these three EPs, we obtain that two EPs are of higher order and are located at $\mathcal{E}=\pm \sqrt{2}t_2$, whereas the last one is a zero-energy EP of order two.

%
%
\begin{figure}[!t]
    \centering
    \includegraphics[width=\columnwidth]{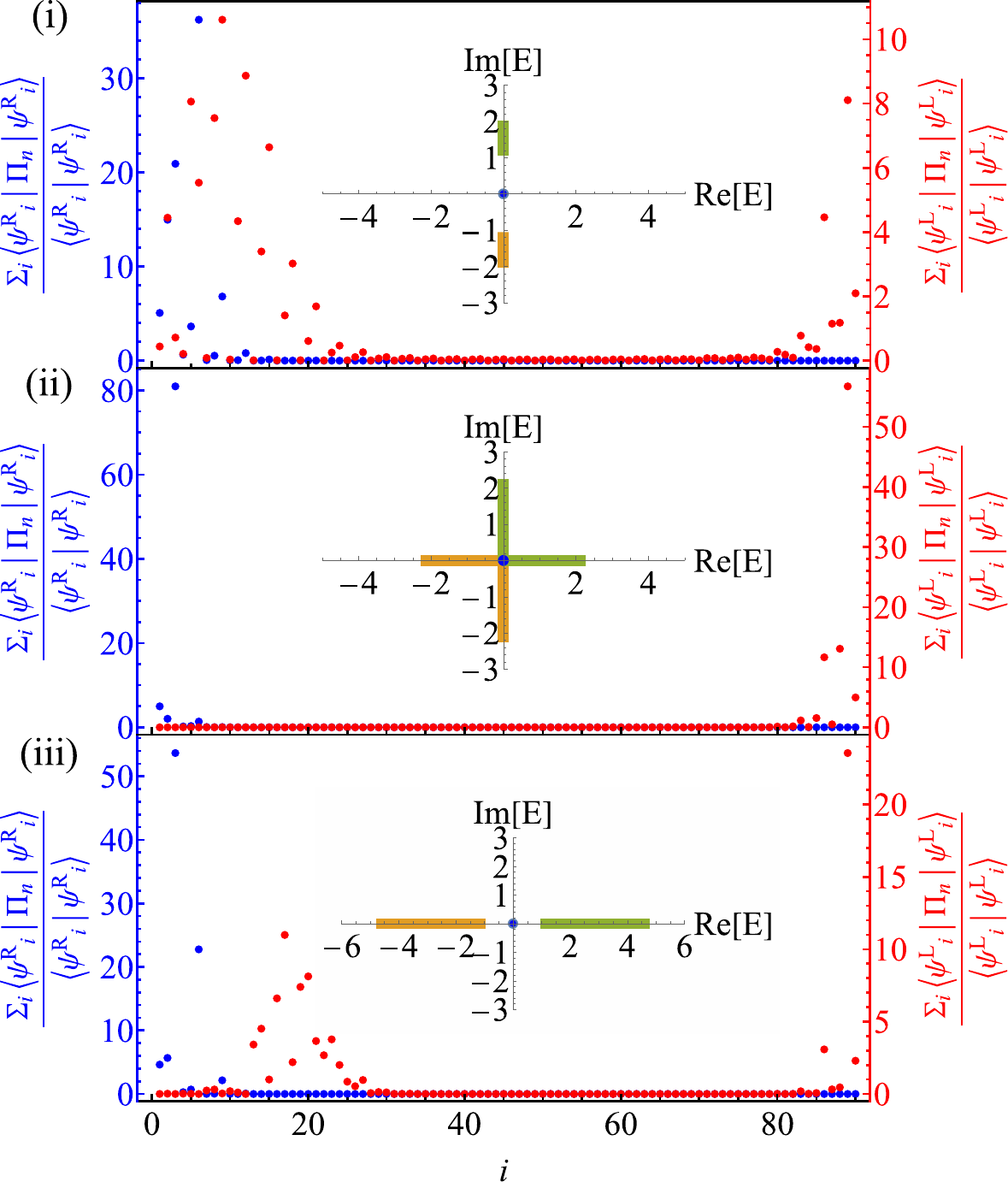}
    \caption{Skin effect of the non-Hermitian DCB chain represented through the sum of the squared amplitudes of each site $i$ with $N=30$ and $\gamma=3$ considering the base $(\mathrm{H}_n,\mathrm{A}_n,\mathrm{B}_n)$. The upper diagram i) represents the sum of the squared amplitudes of each site considering the case in which the eigenvalues are fully imaginary pairs and where also edge states are present, like at $t_1/t_2=0.3$. The middle diagram ii) shows the skin effect under eigenvalues coming in complex conjugate pairs as in $t_1/t_2=1.2$, where the edge states are fading out into the bulk and the lower one iii) represents the case in which the spectra is fully real as it happens for $t_1/t_2=2.7$, with no edge states. Each inset depicts the corresponding positive(green), negative(yellow) and flat(blue) energy bands in the complex plane.}
    \label{fig5}
\end{figure}
%
%
\subsubsection{Diamond Chain B (DCB)}\label{subsubsec:L1B}

For the DCB lattice, the non-Hermitian tight-binding Hamiltonian will read:
%
\begin{align}\label{9} 
\mathcal{H}_{\mathrm{DCB}}=
&\sum_{n} \left\{ \left[t_2(c_{\mathrm{A},n}^{\dagger}+c_{\mathrm{B},n}^{\dagger})c_{\mathrm{H},n+1} 
+\text{h.c.}\right]\right. \nonumber\\ 
&+\left(t_1+\frac{\gamma}{2}\right)(c_{\mathrm{H},n}^{\dagger}c_{\mathrm{A},n}+c_{\mathrm{B},n}^{\dagger}c_{\mathrm{H},n})\\
&+\left.\left(t_1-\frac{\gamma}{2}\right)(c_{\mathrm{A},n}^{\dagger}c_{\mathrm{H},n}+c_{\mathrm{H},n}^{\dagger}c_{\mathrm{B},n})\right\} \nonumber
\end{align}
%
%
in which the primary distinction compared to the DCA lattice is the orientation of the intracell hoppings ($t_1\pm\gamma/2$) connecting the H and B sites | see Fig.~\ref{fig1}b. In the translational invariant form, the corresponding Hamiltonian cannot be expressed anymore only as a function of the characteristic vector $\mathbf{d}$ and the matrices in Eqs.~\eqref{sigmamatrices}, but requires the addition of the $\lambda_7$ Gell-Mann matrix:
%
%
\begin{align}\label{10}
    h_\mathrm{DCB}(\kappa)=h_\mathrm{DCA}(\kappa)-\ii \begin{pmatrix}
        0 & 0 & 0 \\
        0 & 0 & -\ii \\
        0 & \ii & 0 
    \end{pmatrix} \gamma
\end{align}
%
%
This last term is essential in order to change the sign of the non-reciprocal hopping term on the connections between the H and the B sites | see Fig.~\ref{fig1}b.

The corresponding energy spectrum reads   
%
%
\begin{align} \label{spectrumDCB}
    E^{\text{DCB}}_\alpha&= \alpha \sqrt{2 \left(t_1^2+t_2^2\right)+4 t_1 t_2 \cos (\kappa)-\frac{\gamma ^2}{2}}
\end{align}
%
%
with $\alpha\in\{0,\pm\}$. Contrary to the case of DCA in Eq.~\eqref{spectrumDBA}, here we no longer find terms coupling together $\gamma$ with $\kappa$. 
We present in App.~\ref{App1} the analytical expression of the eigenstates of Eq.~\eqref{10}.

Importantly, the expression for the energy spectrum for DCB is, up to a constant factor,  the same as the energy spectrum for the $\mathcal{PT}$-symmetric NH-SSH model~\cite{Lieu_2018}, with the addition of a zero energy flat band. When the magnitudes of the first two terms in Eq.~\eqref{spectrumDCB} exceed that of the third term, a real gap emerges. Conversely, if the first two terms are smaller than the third term, an entirely imaginary gap is present.
When plotted on the complex energy plane, contrary to the case of DCA, the eigenenergies~\eqref{spectrumDCB} collapse into fully imaginary or real lines or  a combination of both | see insets of Fig.~\ref{fig5}.

In spite of the similarities of the energy spectrum for a finite DCB system with a finite $\mathcal{PT}$-symmetric NH-SSH system (see Fig.~\ref{fig4}), the DCB model does not possess EPs under OBC due to its non-Hermitian nature being attributed to non-reciprocal hoppings rather than on-site potentials. As a result, there is no phase transition between a $\mathcal{PT}$-broken and $\mathcal{PT}$-unbroken phase in the DCB model.The complex energy spectrum of the DCB collapsing into lines should signal the disappearance of the skin effect~\cite{Zhang_2022}. However, from Fig.~\ref{fig5}ii,iii, we can clearly observe that the skin effect is still present for the DCB case despite the shape of the energy spectrum in the complex plane. We will give some additional insight into the skin effect of DCB in Sec.~\ref{sec_topo}.

\subsection{Topological properties}\label{sec_topo}

In this section, we are going to explore the topological properties of DCA and DCB. While both exhibit edge modes, DCA possesses zero-energy edge modes, whereas DCB has them at finite energy. Additionally, the flat band eigenstate of the system for the DCB, unlike for DCA, is $k$-dependent | see Eqs.~\eqref{FBBR} and~\eqref{FBBL}. Consequently, we will concentrate on characterizing non-trivial topological properties of the zero-energy modes on the DCA through the biorthogonal polarization~\cite{kunst2018biorthogonal,Bergholtz_2021}. Whereas, for the case of DCB, we will focus on the quantum metric properties of its flat band.

The DCA model exhibits several relevant symmetries considering the 38-fold classification typical on NH systems~\cite{Kawabata_2019}. It presents the Hermitian conjugated particle-hole symmetry (PHS$_c$), with eigenenergies coming as either pure imaginary energies or pairs $(E(k),-E(-k)^*)$. Additionally, it shows the standard time-reversal symmetry (TRS$_c$), where eigenenergies come as real eigenvalues or in complex conjugate pairs, $(E(k),E(-k)^*)$. Moreover, the DCA model possesses the standard sublattice symmetry (SLS), for which the eigenvalues come in $\pm$ pairs at each $k$ point. The DCB model shares the same symmetries as DCA and, in addition, possesses the Hermitian conjugated time-reversal symmetry (TRS$_t$) where the eigenenergies are paired by $(E(k),E(-k))$ and the standard pseudo-Hermiticity symmetry, which restricts the eigenenergies to be real. For a comprehensive overview of the symmetries discussed, please refer to the symmetry table in App.~\ref{App2}.

\subsubsection{Biorthogonal polarization for DCA}
%
%
\begin{figure}[!t]
    \centering
    \includegraphics[width=\columnwidth]{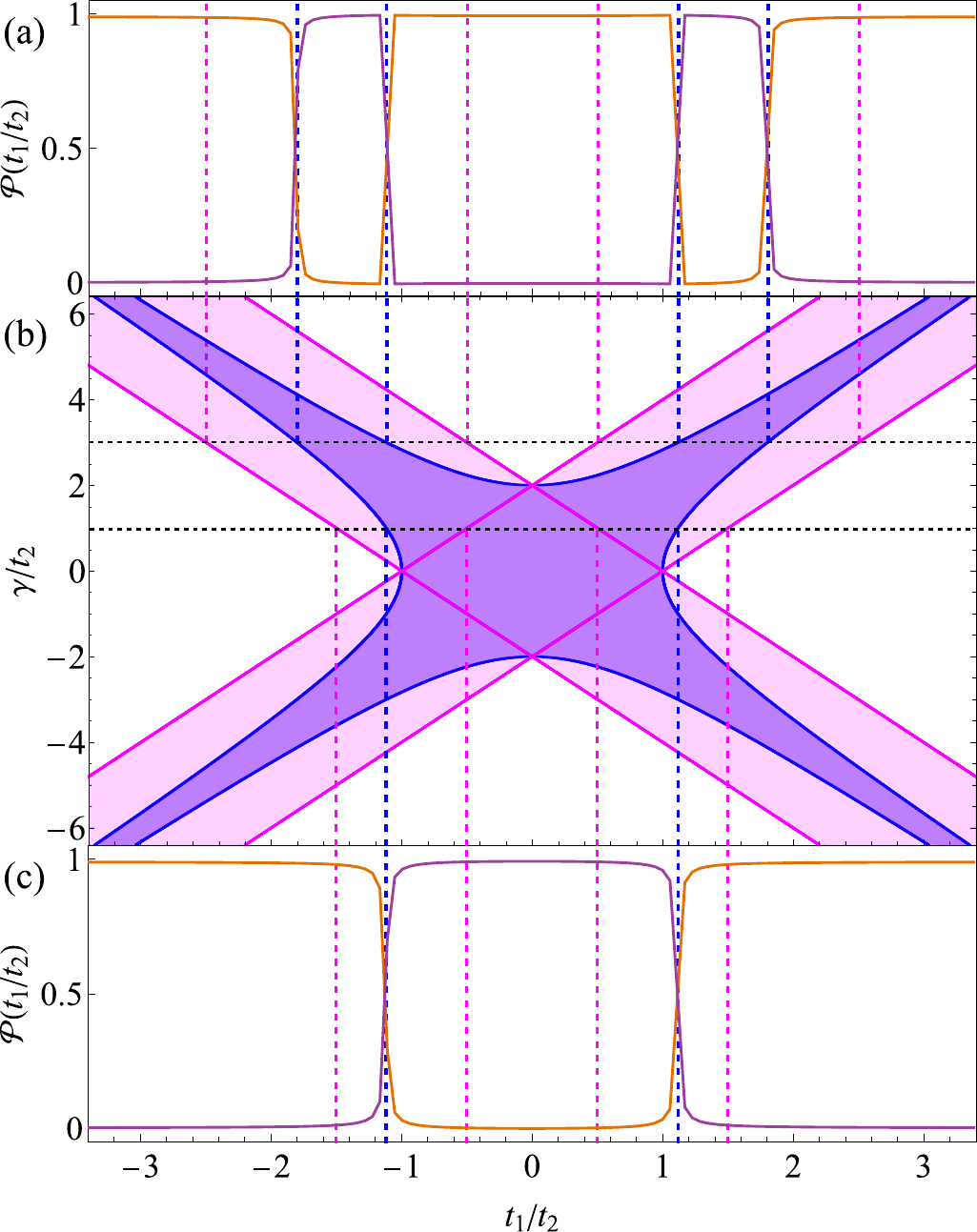}
    \caption{Phase diagram of the DCA model along with the biorthogonal polarization of two cuts in this phase diagram. b) Phase diagram of the non-Hermitian DCA model. In this analysis, pink lines represent the parameter values where exceptional points (EPs) emerge in the eigenenergies under periodic boundary conditions (PBC), while blue lines indicate the appearance or disappearance of zero-energy edge states under open boundary conditions (OBC). In the Hermitian system ($\gamma=0$) under OBC, the phase transition coincides with the one observed in the system under PBC. However, in the non-Hermitian case, the topological phase (purple area) is defined by the OBC system, which broadens the range in which the trivial phase appears (both the pink-shaded and white areas). The dashed lines correspond to the values at which the polarization has been computed analytically for a DCA chain of $N=100$. In panel (a), the calculation is performed for $\gamma/t_2=3$, and in panel (c), it is done at $\gamma/t_2=1$, such that when the zero-energy edge states are present (absent), the biorthogonal polarization is $\mathcal{P}=1(0)$ for the $\mathrm{H}$-broken chain (purple lines) and $\mathcal{P}=0(1)$ for the $\mathrm{AB}$-broken chain (orange lines). The dashed lines in pink and blue indicate the values of the parameters where EPs occur under PBC and phase transition points arise under OBC, respectively.}
    \label{fig6}
\end{figure}
%
%
In this section, we focus on the characterization of the zero-energy boundary modes of the non-Hermitian diamond chain and their localization over the parameter spectrum. We start by combining the left and right non-orthogonal eigenvectors of the non-Hermitian Hamiltonian operators to map the weight distribution of a given band---in this case, the edge-band---over the finite system~\cite{edvardsson2019non}. This arrangement is obtained through the biorthogonal projection expectation value, $\langle\Pi_n\rangle_{\mathrm{LR}}$, also found as the biorthogonal density~\cite{kunst2018biorthogonal} where the biorthogonal projection operator is defined as $ \Pi_n=\sum_\alpha\ket{e_{\mathrm{\alpha},n}}\bra{e_{\mathrm{\alpha},n}}$
with $\ket{e_{\alpha,n}}=c_{\alpha,n}^\dag\ket{0}$~\cite{brody2013biorthogonal}. 
Therefore a real-parameter space topological invariant can be defined, namely the biorthogonal polarization:
%
%
\begin{equation}\label{5.19}
    \mathcal{P}=M-\lim_{N\rightarrow\infty}
    \braket{\Psi^{\mathrm{L}}_0 |N^{-1}\sum_{n=1}^Nn\hat{\Pi}_n|\Psi^{\mathrm{R}}_0}
\end{equation}
%
%
where $M$ is the number of boundary modes and $\ket{\Psi^{\mathrm{R}[\mathrm{L}]}_0}$ are zero-energy right and left boundary modes. The biorthogonal polarization will therefore be $\mathcal{P}=M(0)$ when boundary states are present (absent) at the beginning of a quasi-1D chain.
In order for the wave functions $\ket{\Psi^{\mathrm{R}[\mathrm{L}]}_0}$ in Eq.~\eqref{5.19} to provide significant information for the appearance and
the disappearance of boundary states in the corresponding lattices, the destructive interference method is used. 

This local interference is naturally present on a quasi-1D lattice that begins and also ends with the same motif, in this case with an $\mathrm{H}$ site or with $\mathrm{AB}$ sites, under the constraint of presenting only nearest-neighbor hopping \cite{Kunst1291153, Kunst2019}. Under these circumstances, two chains can be obtained, the so-called broken chains, characterized by an exactly disappearing weight on one of the motifs. As a result, the localization factors of the broken chains' eigenmodes, $r_{\mathrm{L}[\mathrm{R}]}$,  can be exactly solved. These factors, whose magnitude will solely depend on the hopping terms $|r_{\mathrm{L}[\mathrm{R}]}|=f(t_1,t_2,\gamma)$, will eventually give the dispersion rate of the eigenmode, signaling the parameter regions at which the topological phase transitions occur, i.e.,~at which the zero-energy wavefunctions correspond to bulk states $|r^*_{\mathrm{L}}r_{\mathrm{R}}|>1$ or to boundary states $|r^*_{\mathrm{L}}r_{\mathrm{R}}|<1$~\cite{kunst2018biorthogonal}. In spite of the eigenstates of the broken chains not being exact to the ones of the original unbroken chain, the phase transition points will stay unchanged up to finite-size effects in both broken and unbroken lattices~\cite{kunst2018biorthogonal}. 
 
 To further understand the destructive interference method, we will now solve the biorthogonal polarization for the DCA. The number of boundary modes forthis model is $M=1$ for an $\mathrm{H}$-broken chain and $M=1$ for an $\mathrm{AB}$-broken chain. 
 
 For DCA, the zero-energy eigenmodes for the $\mathrm{H}$-broken chain,  $\ket{\Psi^{\mathrm{R}[\mathrm{L}]}_0}_{\mathrm{H}}$, and for the $\mathrm{AB}$-broken chain, $\ket{\Psi^{\mathrm{R}[\mathrm{L}]}_0}_{\mathrm{AB}}$, will be described through
 %
 %
 \begin{widetext}
 \begin{subequations}\label{13}
 \begin{align}
    \ket{\Psi^{\mathrm{R}[\mathrm{L}]}_0}_{\mathrm{H}}&=\mathcal{N}_\mathrm{\mathrm{R}[\mathrm{L}]}\left(1,0,0,r_{\mathrm{R}[\mathrm{L}]},0,0,\cdots,r_{\mathrm{R}[\mathrm{L}]}^{N-2},0,0,r_{\mathrm{R}[\mathrm{L}]}^{N-1},0,0\right) \label{13.a}\\
    \ket{\Psi^{\mathrm{R}[\mathrm{L}]}_0}_{\mathrm{AB}}&=\mathcal{N}_\mathrm{\mathrm{R}[\mathrm{L}]}\left(1,1,0,\frac{1}{r_{\mathrm{R}[\mathrm{L}]}},\frac{1}{r_{\mathrm{R}[\mathrm{L}]}},0,\cdots,\frac{1}{r_{\mathrm{R}[\mathrm{L}]}^{N-2}},\frac{1}{r_{\mathrm{R}[\mathrm{L}]}^{N-2}},0,\frac{1}{r_{\mathrm{R}[\mathrm{L}]}^{N-1}},\frac{1}{r_{\mathrm{R}[\mathrm{L}]}^{N-1}},0\right)
    \label{13.b}
\end{align}
 \end{subequations}
 \end{widetext}
%
%
where we consider a base of $(\mathrm{H}_n,\mathrm{A}_n, \mathrm{B}_n)$ for the $\mathrm{H}$-broken chain and of $(\mathrm{A}_n, \mathrm{B}_n, \mathrm{H}_n)$ for the $\mathrm{AB}$-broken chain to describe each unit cell $n$. In Eqs.~\eqref{13}, $\mathcal{N}_\mathrm{\mathrm{R}[\mathrm{L}]}$ are the right or left normalization factors and $r_{\mathrm{L}[\mathrm{R}]}$ are the left and right localization factors, which are equivalent to those derived for the non-Hermitian SSH model \cite{kunst2018biorthogonal}:
%
%
\begin{equation}
    r_\mathrm{L}=-\frac{t_1+\gamma/2}{t_2} \quad \text{and} \quad  r_\mathrm{R}=-\frac{t_1-\gamma/2}{t_2}
\end{equation}
%
%
The presence of destructive interference in the previous eigenmodes of Eqs.~\eqref{13} can be clearly visualized through the vanishing amplitudes on $\mathrm{A}$ and $\mathrm{B}$ sites for the $\mathrm{H}$-broken chain and in $\mathrm{H}$ sites for the $\mathrm{AB}$-broken one.

Considering the exponential (de)localization of the eigenmodes into the boundaries, $\braket{n|\Psi_0^{\mathrm{R}[\mathrm{L}]}}=\text{e}^{-n/\xi_{\mathrm{R}[\mathrm{L}]}}$ where $\ket{n}=\ket{e_{\mathrm{A},n}}+\ket{e_{\mathrm{H},n}}+\ket{e_{\mathrm{B},n}}$, the jump on the biorthogonal polarization is constructed through the right and left penetration lengths $\xi_{\mathrm{R}[\mathrm{L}]}$ of the boundary modes. Knowing that a biorthogonal
bulk state forms from right and left states localized at
opposite ends $\xi_{\mathrm{R}}=-\xi_{\mathrm{L}}$ and considering that under translational invariance the inverse penetration lengths can be defined through the localization factors as 
%
%
\[\xi^{-1}_{\mathrm{R}[\mathrm{L}]}(\mathbf{k})=\ln|r_{\mathrm{R}[\mathrm{L}]}(\mathbf{k})|\] 
%
%
the condition for having a jump between bulk and boundary eigenmodes will be given by $|r^*_{\mathrm{L}}r_{\mathrm{R}}|=1$~\cite{Kunst1291153}. As a result, the biorthogonal polarization acts as an indicator of the topological phase transition using real-space parameters, $\mathcal{P}=f(t_1,t_2,\gamma)$, resulting in a real-space topological invariant~\cite{edvardsson2019non}. 

Despite the fact that the left and right states might be localized at opposite or the same edges, only one of the boundaries will be considered at a time when analyzing the broken lattices reflecting each end of our unbroken chain. This means that the computation of $\mathcal{P}$ with the left and right eigenstates of Eq.~\eqref{13.a} will represent the localization of boundary states at $n=1$ (see purple lines of Fig.~\ref{fig6}) and the  computation of $M-\mathcal{P}$ through Eq.~\eqref{13.b} will represent the localization of boundary states at $n=N$ (see orange lines of Fig.~\ref{fig6}) for the original chain (starting with an $\mathrm{H}$ site and ending with $\mathrm{A}$ and $\mathrm{B}$ sites).
Considering the shortcut solution used in Ref.~\cite{yao2018edge} for the NH SSH, we obtain the phase transition points of the DCA unbroken chain to be defined as
%
%
\begin{subequations}
    \begin{align}
    t_1&=\pm \sqrt{t_2^2+\left(\frac{\gamma}{2}\right)^2} \quad \text{for} \quad |t_2|>\frac{|\gamma|}{2} \quad \text{and} \\
    t_1&=\pm \sqrt{-t_2^2+\left(\frac{\gamma}{2}\right)^2} \quad \text{for} \quad |t_2|<\frac{|\gamma|}{2}
    \end{align}
\end{subequations}
%
%
which have been represented in blue lines at Fig.~\ref{fig6}b and in blue dashed lines at Fig.~\ref{fig6}a,c. This shows that the appearance of the zero-energy edge states delimited by the biorthogonal polarization is in accordance with the predicted phase transitions from the generalized BZ method used in Ref.~\cite{yao2018edge}.

\subsubsection{Band geometry and topology for DCB}
%
%
\begin{figure}[!t]
    \centering
    \includegraphics[width=0.95\columnwidth]{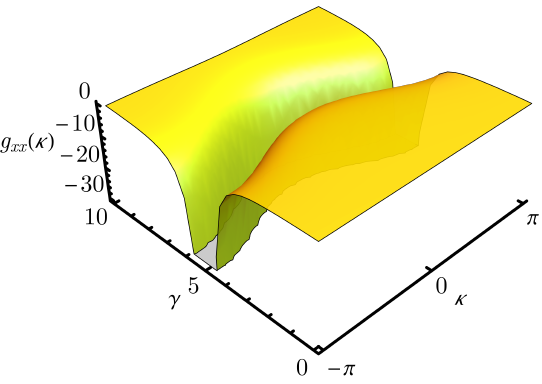}
    \caption{Plot of the quantum metric associated to the flat band~\eqref{metric}of lattice DCB as a function of $t_1$ and of the non-reciprocal coupling term $\gamma$.}
    \label{fig7}
\end{figure}
%
%
As described before, the wave function associated with the flat band of DCB has a non-trivial dependence on the momentum $\kappa$ | see App.~\ref{App1}. In the following, we will show that this leads to non-trivial quantum metric properties. To begin with, we briefly recall the definition and main properties of the quantum metric $g_{\mu\nu}$ in Hermitian and non-Hermitian systems. In the Hermitian case, the quantum metric~\cite{Provost} is a gauge invariant and measurable quantity~\cite{Ozawa2,Cappellaro,Yu} that can be seen as a momentum-space Riemannian metric~\cite{Resta,Palumbo1,Chen,Gritsev}. It has been shown that $g_{\mu\nu}$ plays a central role in Chern insulators~\cite{Roy,Thomale,Ozawa,Northe}, spreading of the Wannier functions~\cite{Marzari}, superconducting weight in flat-band superconductors~\cite{Peotta1,Peotta2,Thumin,Zhu2}, topological semimetals~\cite{Palumbo2,Salerno,Lin,Hwang}, quantum phase transitions~\cite{Mera,Resta2,Chen2} and in the semiclassical equations for wave-packets~\cite{Lapa}. Moreover, the non-Hermitian version of the quantum metric has been originally introduced in Ref.~\cite{DJZhang2019} and has been recently shown to be relevant in different kinds of non-Hermitian topological phases~\cite{Zhu} (see also Ref.~\cite{Smith} for a more recent work on the subject).  In order to define this geometric quantity in the NH case, we first introduce the NH quantum geometric tensor (QGT) $Q_{\mu\nu}^n$, given by
%
%
\begin{align}
Q^n_{\mu\nu}=\frac{1}{2}[\langle \partial_{\mu} u_n^\text{L}|&(1-P_n)|\partial_{\nu} u_n^\text{R}\rangle+ \\ &+\langle\partial_{\mu} u_n^\text{R}|(1-P_n^{\dagger})|\partial_{\nu} u_n^\text{L}\rangle] \nonumber
\end{align}
%
%
where $n$ is the band index, $|u_n^\text{L/R}\rangle$ are the left/right Bloch wave-eigenvectors, $\partial_{\mu}=\partial_{k_{\mu}}$ and $P_n=|u_n^\text{R}\rangle\langle u_n^\text{L}|$ is the projector operator.
Thus, the corresponding NH quantum metric tensor is given by the real part of the above tensor, i.e., $g_{\mu\nu}=\text{Re}(Q_{\mu\nu}^n)$ while the imaginary part of the NH QGT corresponds to the NH Berry curvature, namely $F^n_{\mu\nu}=-2\text{Im}(Q^n_{\mu\nu})$. However, we notice that in 1D systems, the imaginary part of the QGT is absent such that $g_{\mu\nu}$ is the only non-trivial geometric quantity that can be built. For the DCB, the NH quantum metric for the flat and negative energy bands are respectively given by
%
%
\begin{align}\label{metric}
g^{0}_{xx}(\kappa)&=-\frac{ t_2^2\, \gamma^2}{\left(E^{\mathrm{DCB}}_-\right)^4}  \\
g^{-}_{xx}(\kappa)&= -\frac{ t_2^2\, (\gamma^2-t_1^2 - 2 t_2^2 - 4 t_1 t_2 \cos \kappa - t_1^2 \cos 2\kappa)}{2\left(E^{\mathrm{DCB}}_-\right)^4}\nonumber
\end{align}
%
%
We can now make some relevant observations concerning the peculiar band geometry of DCB: first, $g^{0}_{xx}$ is non-zero only in the NH regime, in which even the flat band acquires some non-trivial geometric features differently from the Hermitian case. In fact, due to its dependence on the energy dispersion, $E^{\mathrm{DCB}}$, the band geometry of the flat band contains information about the 
exceptional points of DCB. Thus the corresponding quantum metric is divergent at those points, similar to what has been shown in NH models without flat bands~\cite{Malpuech1,Malpuech2}. 
Second, another main consequence of the non-trivial behavior of $g^{0}_{xx}$ is related to the conductivity
at the flat band that is robust in the presence of disorder and can even be tremendously boosted, similar to the Hermitian case discussed in Ref.~\cite{Bouzerar_2022}. The boost is obtained for values of $\gamma$ changing the spectrum from real to imaginary~\eqref{spectrumDCB} | see Fig.~\ref{fig7}. Third, in the limit $t_1 \rightarrow 0$, DCB supports three completely flat bands, and the quantum metric becomes constant for all the bands. This quantum geometry behavior resembles the one related to the Hermitian two-band Creutz ladder, where the quantum metric has been shown to be constant for both flat bands~\cite{Liang}.
In this limit, the ratio between $g^{0}_{xx}$ and $g^{-}_{xx}$ acquires a simple expression
%
%
\begin{equation}
\frac{g^{0}_{xx}}{g^{-}_{xx}}=\frac{2\gamma^2}{\gamma^2-2 t_2^2}
\end{equation}
%
%
which converges to 2 for $\gamma \gg t_2$, i.e., in the strongly NH regime. On the other hand, in the opposite Hermitian limit $\gamma=0$ and for $t_1 \rightarrow 0$, ${g^{-}_{xx}}$ is still constant and equal to $1/4$ while 
$g^{0}_{xx}$ becomes identically null.

\noindent To characterize the band topology of DCB, we employ the approach originally proposed in the Hermitian framework in Ref.~\cite{Palumbo3} and then extended in the non-Hermitian systems in Ref.~\cite{Zhu}, which is based on the construction of suitable momentum-space scalar fields $\phi^\text{L}_n$ and $\phi^\text{R}_n$, given by
%
%
\begin{equation}
\phi^\text{L}_n=-\frac{\ii}{2}\log\prod_{\aleph} u^\text{L}_{n,\aleph}, \hspace{0.3cm} \phi^\text{R}_n=-\frac{\ii}{2}\log\prod_{\aleph} u^\text{R}_{n,\aleph}
\end{equation}
%
%
where $u^\text{L}_{n,\aleph}$ and $u^\text{R}_{n,\aleph}$ denote
the non-zero components of $|u^\text{L}_{n}\rangle$ and $|u^\text{R}_{n}\rangle$, respectively. 
From the above scalar fields in our 1D case, we can define a NH Zak-like phase for the DCB as follows
%
%
\begin{equation}
w_n= -\int_0^{2\pi}d\kappa\, \partial_{\kappa} \phi^\text{L}_n
=-\int_0^{2\pi}d\kappa\, \partial_{\kappa} \phi^\text{R}_n
\end{equation}
%
%
where the integration is performed in the first Brillouin zone. From this expression, we observe that for the flat band, $w_0$ is never quantized, which implies a trivial band topology. On the other hand, the lower band carries non-trivial band topology in the following regime
%
%
\begin{equation}
w_-= \pi, \hspace{0.3cm} |t_2|-|t_1|>|\gamma|/2
\end{equation}
%
%
This result represents a natural but non-trivial generalization of the topological phase supported for $|t_2|>|t_1|$  in the Hermitian model~\cite{Bercioux_2017}.

%
%
\begin{figure}[!t]
    \centering    \includegraphics[width=0.8\columnwidth]{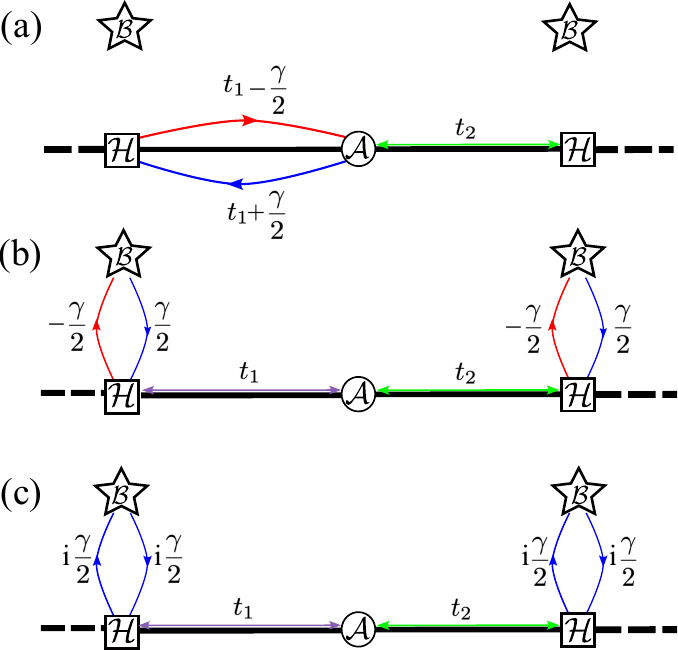}    
    \caption{Rotated diamond chain from the DCA to a) a non-Hermitian SSH chain with non-reciprocal hoppings plus disconnected $\mathcal{B}$ sites and from the DCB to a Hermitian SSH chain coupled through b) non-reciprocal hopping parameters to $\mathcal{B}$ sites using $U_1$ matrix transformation or c) complex hopping parameters  to the virtual sites using $U_\text{i}$, respectively.}
    \label{fig8}
\end{figure}
%
%
%
%
\begin{figure}[!t]
    \centering
    \includegraphics[width=0.9\columnwidth]{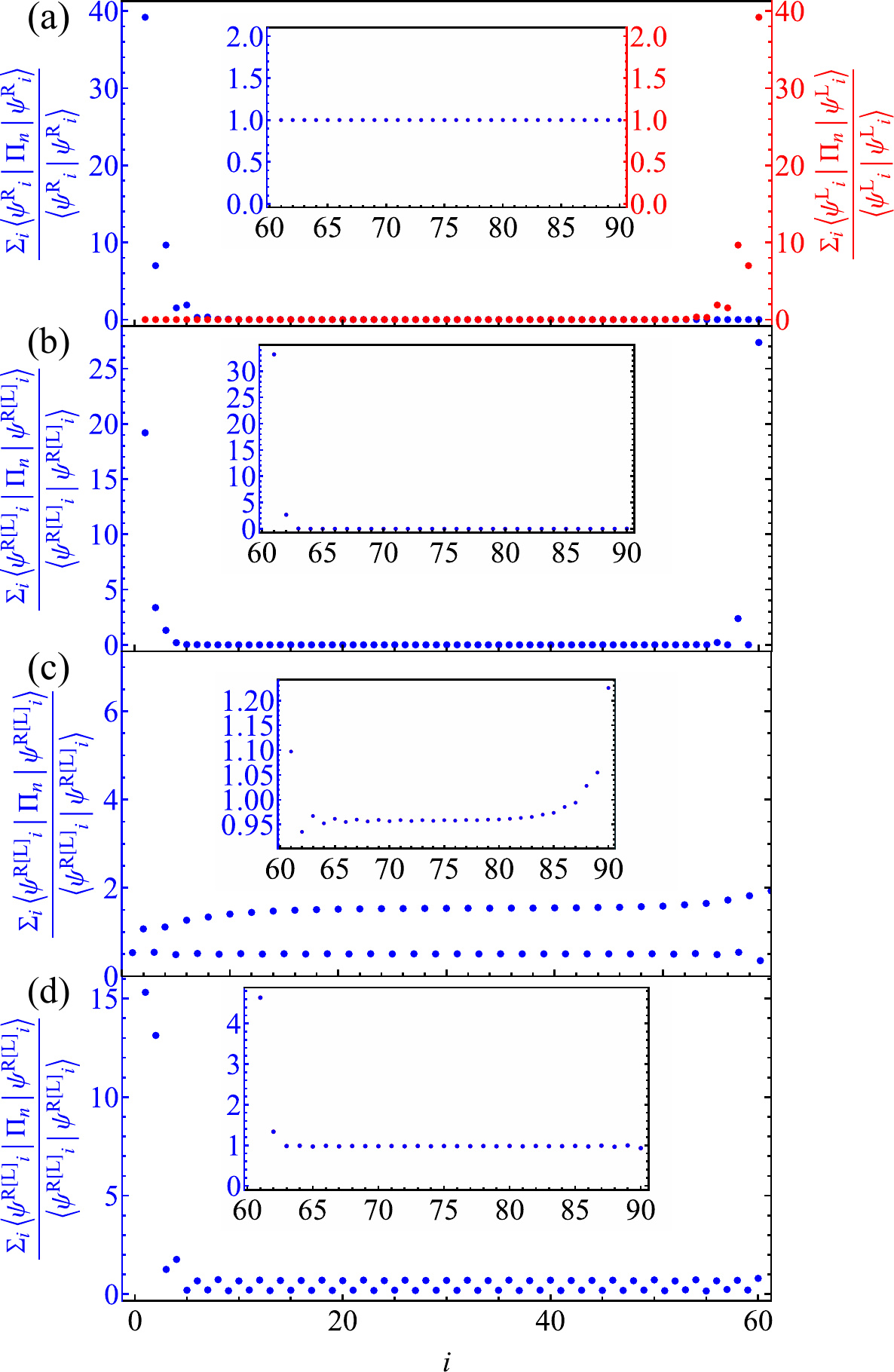}
    \caption{Skin effect for the rotated non-Hermitian DCA and DCB chains represented through the sum of the squared amplitudes of each site $i$ with $N=30$ and $\gamma=3$. The upper diagram (a) represents the skin effect of the rotated non-Hermitian DCA system at $t_1/t_2=2$, the second panel (b), shows the first case of the rotated non-Hermitian DCB system where the eigenvalues are fully imaginary pairs like at $t_1/t_2=0.3$, the third diagram (c) shows the skin effect of the rotated non-Hermitian DCB model skin effect under eigenvalues coming in complex conjugate pairs as in $t_1/t_2=1.2$ and the lower one (d) represents the rotated non-Hermitian DCB skin effect for the case in which the spectra is fully real as it happens for $t_1/t_2=2.7$. In the present figure, 
    the rotated systems have been represented considering the base $(\mathcal{H}_n,\mathcal{A}_n,\mathcal{B}_n)$. Additionally, the sites of the open chain have been redistributed such that the skin effect of the corresponding $\mathcal{B}$ sites are depicted in the inset of each figure, leaving the main figure for the skin effect originated from the $\mathcal{A}$ and $\mathcal{H}$ sites, respectively.}
    \label{fig9}
\end{figure}
%
%
\subsection{Rotation into a 1D model with a flat band}\label{sec_rot}

In the following, we show how to perform a rotation to a new base choice of the diamond chain, which allows us to have a better understanding of several of the properties we have shown so far. We obtain the new base considering a real or complex linear combination of the site with lower connectivity. The site operators in this new base read
%
%
\begin{subequations}\label{rotation}
\begin{align}
\mathcal{A}_n^\beta&=\frac{1}{\sqrt{2}}(c_{\mathrm{A},n}+\beta c_{\mathrm{B},n}) \\
\mathcal{H}_n&=c_{\mathrm{H},n} \\ \mathcal{B}_n^\beta&=\frac{1}{\sqrt{2}}(c_{\mathrm{A},n}-\beta c_{\mathrm{B},n})
\end{align}
\end{subequations}
%
%
where $\beta=\{1,\ii\}$. The matrices $U_\beta$ associated with this rotation result in:
%
%
\begin{align}\label{20}
  U_\mathrm{\beta}=\frac{1}{\sqrt{2}}
  \begin{pmatrix}
   1 & 0 & \beta \\
   0 & \sqrt{2} & 0 \\
   1 & 0 & -\beta \\
 \end{pmatrix}
\end{align}
%
%

and applied to the NH Hamiltonian operator in reciprocal space of Eq.~\eqref{2}, give rise to the same rotated model of the DCA. The resulting Hamiltonian in reciprocal space reads:
%
%
\begin{equation}\label{21}
h_\mathrm{DCA}^{\mathrm{Rot}}(\kappa)=d_x\Sigma_x^{\mathrm{Rot}}+d_y\Sigma_y^{\mathrm{Rot}}
\end{equation}
%
%
being the $\Sigma_x^{\mathrm{Rot}}$ and $\Sigma_y^{\mathrm{Rot}}$ matrices defined as:
%
%
\label{newsigmamatrices}
\begin{align}\label{22}
    \Sigma_x^{\mathrm{Rot}}=
     \begin{pmatrix}
        0 & 1 & 0 \\
        1 & 0 & 0 \\
        0 & 0 & 0 \\
    \end{pmatrix}
    ~ \text{and} ~ \Sigma_y^{\mathrm{Rot}}=
    \begin{pmatrix}
        0 & \text{i} & 0 \\
        -\text{i} & 0 & 0 \\
        0 & 0 & 0 \\
    \end{pmatrix}
\end{align}
%
%
which is analogous to the non-Hermitian Stub model of Ref.~\cite{Bartlett_2021} with $t_3=0$. The rotated DCA model, therefore, describes a non-Hermitian SSH chain with non-reciprocal hopping parameters [see Fig.~\ref{fig8}a] formed by $\mathcal{A}$ and H sites decoupled to the $\mathcal{B}$ sites originating the flat band. This rotation helps to understand that in this new base, the NH skin effect and the topological properties should be analogous to the one of the NH SSH model~\cite{kunst2018biorthogonal}.

The transformation matrices defined in Eq.~\eqref{20} applied to the DCB lattice in reciprocal space described in Eq.~\eqref{10} give rise to two equivalent systems for real and complex $\beta$: 
%
%
\begin{subequations}\label{DCB_rotated}
    \begin{align}
    h_{\mathrm{DCB}}^{\mathrm{Rot1}}(\kappa)&=U_1^{-1}h_{\mathrm{DCB}}(\kappa) U_1 \label{14} \nonumber \\ & = \sqrt{2}
  \begin{pmatrix}
            0 & t_1+t_2\text{e}^{\text{i}\kappa} & 0\\
          t_1+t_2\text{e}^{-\text{i}\kappa} & 0 & \frac{\gamma}{2}\\
            0 & -\frac{\gamma}{2} & 0
          \end{pmatrix}\\
    h_{\mathrm{DCB}}^{\mathrm{Rot2}}(\kappa)&=U_\ii^{-1} h_{\mathrm{DCB}}(\kappa) U_\ii \label{15} \nonumber \\  &= \sqrt{2}
  \begin{pmatrix}
            0 & t_1+t_2\text{e}^{\text{i}\kappa} & 0\\
          t_1+t_2\text{e}^{-\text{i}\kappa} & 0 & \text{i}\frac{\gamma}{2}\\
            0 & \text{i}\frac{\gamma}{2} & 0
          \end{pmatrix}
\end{align}
\end{subequations}
%
%

Both of these models contain a Hermitian SSH chain formed by $\mathcal{A}$ and $\mathcal{H}$ sites that are now coupled to the $\mathcal{B}$ sites in a non-Hermitian fashion, cf. Figs.~\ref{fig8}b,c. It is important to note that the $\mathcal{B}$ sites are disconnected from each other, thus representing the flat band. However, they are coupled to the Hermitian SSH model. This coupling is found to be non-reciprocal for the first rotation in Eq.~\eqref{14}, where we observe that we have broken the inversion symmetry $\mathcal{I}=\Sigma_x$ | see Fig.~\ref{fig8}b. While the case of the imaginary rotation in  Eq.~\eqref{15}, the coupling between the flat band and the Hermitian SSH chain breaks time-reversal symmetry $\mathcal{T}=\mathcal{C}$ | see Fig.~\ref{fig8}c. This rotation helps us to unveil the nature of the NH skin effect we have presented in Fig.~\ref{fig5}. The dispersive part of the system is a Hermitian SSH model and should not present any skin effect; however, it is connected in a NH fashion to the flat band giving rise to a finite skin effect. It has to be noted that the strength of this skin effect is smaller than in the case of DCA since it originates in the flat band where all the $\mathcal{B}$ sites are disconnected. 

In Fig.~\ref{fig9}, we present the results for the skin effect for DCA and DCB, considering the tight-binding representations of Eqs.~\eqref{21} and~\eqref{DCB_rotated}. For DCA in Fig.~\ref{fig9}a,
 we can clearly observe a skin effect identical to the case of the NH SSH for the $\mathcal{H}$ and $\mathcal{A}$, whereas, for the $\mathcal{B}$ sites corresponding to the flat band, the skin effect is absent. For the case of DCB, we show in Fig.~\ref{fig9}b-d the results for the same parameter values we used in Fig.~\ref{fig5} but in the rotated base, i.e., $\{\mathcal{H},\mathcal{A},\mathcal{B}\}$. We clearly observe that now the skin effect that arises in the flat band is transmitted by \emph{proximity} to the Hermitian SSH model via the NH coupling. Most importantly, within this choice of the base, the right,
and the left eigenvectors differ only up to a phase factor, for the reason they are identical when representing them in Fig.~\ref{fig9}b,d.

\section{Discussions and Conclusion}\label{Conclusions}

In this article, we have investigated the topological properties of a non-Hermitian version of the diamond chain. We have shown that compared to the Hermitian case, there exist two possible non-Hermitian configurations presenting non-trivial topological properties. Specifically, for one of the two configurations that we have named DCA, the system presents properties analogous to a non-Hermitian SSH chain plus a flat band. We have characterized this system by investigating the biorthogonal polarization. The second configuration, which we have named DCB, presents a non-zero quantum metric for the flat band due to the wave function's $k$ dependency. This quantum metric becomes highly divergent near the exceptional points, providing a significant boost when tuning the non-Hermitian parameter $\gamma$.  
The band topology of the DCB configuration indicates that the flat band possesses a trivial nature, while the non-Hermitian Zak phase of the lower band can exhibit a non-trivial band topology. Both systems can be mapped to the SSH model. The DCA system can be transformed into a non-Hermitian SSH model with disconnected sites representing the flat band. In contrast, the DCB model can be similarly transformed into a Hermitian SSH chain, connected in a non-Hermitian fashion to these additional sites representing the flat band. Interestingly, both systems present a NH skin effect. This is especially surprising for DCB, which presents a spectrum in the complex plane that is either real, imaginary, or a combination of both.

\begin{acknowledgments}
The authors acknowledged useful discussions with Duy Hoang Minh Nguyen and Alejandro Sebasti\'an G\'omez G\'omez. 
C.M.S, M.A.J.H, and D.B.  acknowledged the support from the Spanish MICINN-AEI through Project No. PID2020-120614GB-I00~(ENACT). D.B. further acknowledged the support from  Transnational Common Laboratory $Quantum-ChemPhys$, and the Department of Education of the Basque Government through
the project PIBA\_2023\_1\_0007 (STRAINER). A.G.E., D.B., M.A.J.H., and C.M. S. acknowledge funding from the IKUR Strategy under the collaboration agreement between Ikerbasque Foundation and DIPC on behalf of the Department of Education of the Basque Government, Programa de ayudas de apoyo a los agentes de la Red Vasca de Ciencia, Tecnolog\'ia e Innovaci\'on acreditados en la categor\'ia de Centros de Investigaci\'on B\'asica y de Excelencia (Programa BERC) from the Departamento de Universidades e Investigaci\'on del Gobierno Vasco and Centros Severo Ochoa AEI/CEX2018-000867-S from the Spanish Ministerio de Ciencia e Innovaci\'on and from the Gipuzkoa Provincial Council within the QUAN-000021-01 project. A.G.E., M.A.J.H., and C.M.S. acknowledged support from the Spanish Ministerio de Ciencia e Innovaci\'on (PID2022-142008NB-I00) and the Basque Government Elkartek program (KK-2023/00016). F.K.K. acknowledges funding from the European Union via the ERC Starting Grant ``NTopQuant''. Views and opinions expressed are, however, those of the authors only and do not necessarily reflect those of the European Union or the European Research Council (ERC). Neither the European Union nor the ERC can be held responsible for them.
\end{acknowledgments}


\section*{Appendix A}
\renewcommand{\thesubsection}{A.\arabic{subsection}} %
\setcounter{equation}{0}
\renewcommand\theequation{A\arabic{equation}}

\subsection{Eigenvectors of the non-Hermitian diamond chain}\label{App1}
In this appendix, we will present the analytical expressions for the eigenvectors of the two lattice systems, DCA and DCB. We start by introducing the following quantities:
%
%
\begin{align}
    \rho_\pm(\kappa)&= \frac{t_1\pm\frac{\gamma}{2}}{t_2}+\cos\kappa +\ii \sin\kappa \nonumber \\
    &=|\rho_\pm(\kappa)|\ee^{\ii \phi_\pm(\kappa)}
\end{align}
%
%
The biorthogonal base for DCA reads:
%
%
\begin{subequations}\label{Right_DCA}
   \begin{align}
    |\psi_{0,\text{R}}\rangle &= \frac{1}{\sqrt{2}} \left( -1,0,1\right)
    \label{FBAR} \\
    |\psi_{\alpha,\text{R}}\rangle&= \frac{1}{2}\left(1 ,\frac{E_\alpha^\text{DCA}}{t_2 |\rho_-(\kappa)|\ee^{\ii \phi_-}},1\right) \label{DBAR}
\end{align}
\end{subequations}
%
%
and 
%
%
\begin{subequations}\label{Left_DCA}
\begin{align}
|\psi_{0,\text{L}}\rangle &= \frac{1}{\sqrt{2}} \left( -1,0,1\right) \label{FBAL} \\
|\psi_{\alpha,\text{L}}\rangle&= \frac{1}{2}\left(1 ,\frac{E_\alpha^\text{DCA}
}{t_2 |\rho_+(\kappa)|\ee^{\ii \phi_+}},1\right) \label{DBAL}
\end{align}
\end{subequations}
%
%
These states fulfill the biorthogonal scalar product~\cite{brody2013biorthogonal}:
%
%
\begin{align}\label{Biorthogonal}
\langle \psi_{\alpha,\beta,\text{R}}|\psi_{\alpha',\beta',\text{L}}\rangle=\delta_{\alpha,\alpha'}\delta_{\beta,\beta'}
\end{align}
%
%
Whereas the biorthogonal base for DCB reads:
%
%
\begin{subequations}\label{Right_DCB}
\begin{align}
|\psi_{0,\text{R}}\rangle &= \mathcal{N} \left( -|\rho_+(\kappa)| \ee^{-\ii \phi_+} + \gamma,0,|\rho_+(\kappa)| \ee^{\ii \phi_+}\right), \label{FBBR}\\
|\psi_{\alpha,\text{R}}\rangle&= \mathcal{N}\left(\frac{-t_2|\rho_-(\kappa)| \ee^{-\ii \phi_-} + \gamma}{t_2 |\rho_+(\kappa)| \ee^{\ii \phi_+}},\frac{E_\alpha^\text{DCB}}{t_2 |\rho_+(\kappa)| \ee^{\ii \phi_+}},1\right) \label{DBBR}
\end{align}
\end{subequations}
%
%
and 
%
%
\begin{subequations}\label{Left_DCB}
\begin{align}
|\psi_{0,\text{L}}\rangle &= \mathcal{N} \left( -|\rho_+(\kappa)|\ee^{-\ii \phi_+}|,0,|\rho_-(\kappa)|\ee^{\ii \phi_-}|\right) \label{FBBL}\\
|\psi_{\alpha,\text{L}}\rangle&= \mathcal{N}\left(\frac{t_2|\rho_+(\kappa)| \ee^{\ii \phi_+} + \gamma}{t_2|\rho_-(\kappa)| \ee^{\ii \phi_-}},\frac{E_\alpha^\text{DCB}}{t_2 |\rho_-(\kappa)| \ee^{\ii \phi_-}},1\right)  \label{DBBL}
\end{align}
\end{subequations}
%
%
Here, we have introduced the factor of normalization $\mathcal{N}$ in order to fulfill the biorthogonal scalar product in Eq.~\eqref{Biorthogonal}; the normalization factor is defined as
%
%
\begin{equation}
    \mathcal{N}=\sqrt{\frac{1}{2}+\frac{2 \ii \gamma  t_2 \sin\kappa}{4\left(t_1^2+t_2^2\right)-\gamma ^2+8 t_1 t_2 \cos\kappa
   }}
\end{equation}
%
%

%
%
\begin{table}[!t]
 \centering
 \caption{The DCA model is subject to symmetry conditions, each with its specific constraint equation. These equations are fulfilled by applying the symmetry operator on the Hamiltonian, where $\beta=\{1,\text{i}\}$.}
  \label{table:symmetry-conditions_DCA}
 \begin{tabular}{llll}
    \hline
    Symmetry & Equation & Operator\\
    \hline
    PHS$_t$, $CC^* =  \mathbf{1}$ & $h(-k) = -Ch^T(k)C^{-1}$ & $-$ \\
    PHS$_t$, $CC^* = - \mathbf{1}$ & $h(-k) = -Ch^T(k)C^{-1}$ & $-$   \\
    TRS$_t$, $TT^* =  \mathbf{1}$ & $h(-k) = Th^{T}(k)T^{-1}$ &  $-$   \\
    TRS$_t$, $TT^* = - \mathbf{1}$ & $h(-k) = Ch^T(k)C^{-1}$ & $-$   \\
    PHS$_c$, $CC^* =  \mathbf{1}$ & $h(-k) = -Ch^*(k)C^{-1}$ &  $\beta R_{\mathrm{A},\mathrm{B}}, \beta R_{\mathrm{H}}$,\\
     & & $\beta G_{\mathrm{A},\mathrm{B}}, \beta G_{\mathrm{H}}$   \\
    PHS$_c$, $CC^* = - \mathbf{1}$ & $h(-k) = -Ch^*(k)C^{-1}$ & $-$   \\
    TRS$_c$, $TT^* =  \mathbf{1}$ & $h(-k) = Th^*(k)T^{-1}$ & $  \beta\mathbb{I}_{3\times3}, \beta P_{\mathrm{B},\mathrm{A}}$   \\
    TRS$_c$, $TT^* = - \mathbf{1}$ & $h(-k) = Th^*(k)T^{-1}$ & $-$   \\
    CS, $\Gamma^2 =  \mathbf{1}$ & $h(k) = -\Gamma h^\dagger(k)\Gamma^{-1}$ & $-$   \\
  
    Pseudo-Hermiticity, \\
    $\eta^2 =  \mathbf{1}$  & $h(k) = \eta h^\dagger(k)\eta^{-1}$ & $-$   \\
    SLS, $S^2 =  \mathbf{1}$ & $h(k) = -Sh(k)S^{-1}$ &  $R_{\mathrm{A},\mathrm{B}}, R_{\mathrm{H}}$,\\
     & & $ G_{\mathrm{A},\mathrm{B}}, G_{\mathrm{H}}$    \\
    Parity, $P^2 =  \mathbf{1}$ & $h(-k) = Ph(k)P^{-1}$ & $-$   \\
    Parity-time, \\ 
    $(PT)(PT)^* =  \mathbf{1}$ & $h(k) = (PT)h^*(k)(PT)^{-1}$ & $-$   \\
     Parity-time, \\
     $(PT)(PT)^* = - \mathbf{1}$ & $h(k) = (PT)h^*(k)(PT)^{-1}$ & $-$  \\

    \hline
  \end{tabular}
\end{table}
%
%

%
%
\begin{table}[!t]
  \centering
 \caption{The DCB model is subject to symmetry conditions, each with its specific constraint equation. These equations are fulfilled by applying the symmetry operator on the Hamiltonian, where $\beta=\{1,\text{i}\}$.}
  \label{table:symmetry-conditions_DCB}
 \begin{tabular}{llll}
    \hline
    Symmetry & Equation & Operator\\
    \hline
    PHS$_t$, $CC^* =  \mathbf{1}$ & $h(-k) = -Ch^T(k)C^{-1}$ & $-$ \\
    PHS$_t$, $CC^* = - \mathbf{1}$ & $h(-k) = -Ch^T(k)C^{-1}$ & $\beta G_{\mathrm{A},\mathrm{B}}, \beta G_{\mathrm{H}}$    \\
    TRS$_t$, $TT^* =  \mathbf{1}$ & $h(-k) = Th^{T}(k)T^{-1}$ &  $\beta P_{\mathrm{B},\mathrm{A}}$  \\
    TRS$_t$, $TT^* = - \mathbf{1}$ & $h(-k) = Ch^T(k)C^{-1}$ & $-$   \\
    PHS$_c$, $CC^* =  \mathbf{1}$ & $h(-k) = -Ch^*(k)C^{-1}$ &  $\beta R_{\mathrm{A},\mathrm{B}}, \beta R_{\mathrm{H}}$   \\
    PHS$_c$, $CC^* = - \mathbf{1}$ & $h(-k) = -Ch^*(k)C^{-1}$ & $-$   \\
    TRS$_c$, $TT^* =  \mathbf{1}$ & $h(-k) = Th^*(k)T^{-1}$ & $  \beta\mathbb{I}_{3\times3}$   \\
    TRS$_c$, $TT^* = - \mathbf{1}$ & $h(-k) = Th^*(k)T^{-1}$ & $-$   \\
    CS, $\Gamma^2 =  \mathbf{1}$ & $h(k) = -\Gamma h^\dagger(k)\Gamma^{-1}$ & $G_{\mathrm{A},\mathrm{B}}, G_{\mathrm{H}}$    \\
    Pseudo-Hermiticity, \\
    $\eta^2 =  \mathbf{1}$ & $h(k) = \eta h^\dagger(k)\eta^{-1}$ & $P_{\mathrm{B},\mathrm{A}}$   \\
    SLS, $S^2 =  \mathbf{1}$ & $h(k) = -Sh(k)S^{-1}$ &  $R_{\mathrm{A},\mathrm{B}}, R_{\mathrm{H}}$   \\
    Parity, $P^2 =  \mathbf{1}$ & $h(-k) = Ph(k)P^{-1}$ & $-$   \\
    Parity-time, \\
    $(PT)(PT)^* =  \mathbf{1}$ & $h(k) = (PT)h^*(k)(PT)^{-1}$ & $-$   \\
     Parity-time, \\
     $(PT)(PT)^* = - \mathbf{1}$ & $h(k) = (PT)h^*(k)(PT)^{-1}$ & $-$  \\
    \hline
  \end{tabular}
\end{table}

\subsection{Table of symmetries}\label{App2}

In this appendix, we present the table on the symmetries of DBA and DCB. We introduce the following auxiliary matrices:
%
%
\begin{subequations}
\begin{align}
    R_{\mathrm{H}}&=\begin{pmatrix}
      -1  & 0 & 0 \\
       0  & 1 & 0 \\
       0  & 0 & -1 
   \end{pmatrix} \\
R_{\mathrm{A},\mathrm{B}}&=\begin{pmatrix}
      1  & 0 & 0 \\
       0  & -1 & 0 \\
       0  & 0 & 1 
    \end{pmatrix} \\
    P_{\mathrm{B},\mathrm{A}}&=\begin{pmatrix}
      0  & 0 & 1 \\
       0  & 1 & 0 \\
       1  & 0 & 0 
    \end{pmatrix}\\
G_{\mathrm{H}}&=R_{\mathrm{H}}P_{\mathrm{B},\mathrm{A}}, \quad \text{and}\\
    G_{\mathrm{A},\mathrm{B}}&=R_{\mathrm{A},\mathrm{B}}P_{\mathrm{B},\mathrm{A}},
\end{align} 
\end{subequations}
%
%
which represent different symmetry operations. Let's consider the following matrices representing a simplified version of the  Hamiltonians of the DCA and DCB models in reciprocal space:
%
%
 \begin{align}
 \label{simpleham}
    h_{\mathrm{DC}}= 
      \begin{pmatrix}
       0 & t_{\mathrm{A}\rightarrow\mathrm{H}} & 0\\
       t_{\mathrm{H}\rightarrow\mathrm{A}} & 0 & t_{\mathrm{H}\rightarrow\mathrm{B}}\\
        0 & t_{\mathrm{B}\rightarrow\mathrm{H}} & 0
      \end{pmatrix}
\end{align}
%
%
being $h_{\mathrm{DCB}}$ like $h_{\mathrm{DCA}}$ but changing $t\rightarrow t'$. The matrix $R_\mathrm{H}$ when applied at the left of the Hamiltonians described in Eq.~\eqref{simpleham} changes the sign of the $\mathrm{H}$-outgoing hoppings, $t^{(')}_{\mathrm{H}\rightarrow\mathrm{A,B}}\Rightarrow -t^{(')}_{\mathrm{H}\rightarrow\mathrm{A,B}}$. Similarly, the matrix $R_\mathrm{A,B}$, when applied at the left of the Hamiltonians, changes the sign of the $\mathrm{A}$-- and  $\mathrm{B}$--outgoing hoppings, $t^{(')}_{\mathrm{A,B}\rightarrow \mathrm{H}}\Rightarrow -t^{(')}_{\mathrm{A,B}\rightarrow\mathrm{H}}$. Consequently, when the full transformations are applied, the resulting matrix is a complete sign flip for both transformations: $h\Rightarrow R_\mathrm{H} h R^{-1}_\mathrm{H}=-h$ and $h \Rightarrow R_\mathrm{A,B}hR^{-1}_\mathrm{A,B}=-h$.

\pagebreak

The permutation matrix $P_\mathrm{B,A}$ exchanges the hopping parameters between the $\mathrm{A}$ and $\mathrm{B}$ sites $t^{(')}_{\mathrm{H}\rightarrow\mathrm{A,B}}\Rightarrow t^{(')}_{\mathrm{H}\rightarrow\mathrm{B,A}}$ and $ t^{(')}_{\mathrm{A,B}\rightarrow\mathrm{H}}\Rightarrow t^{(')}_{\mathrm{B,A}\rightarrow\mathrm{H}}$. Consequently, the transformation matrices $G_\mathrm{H}$ and $G_\mathrm{A,B}$ combine both transformations, resulting in an overall sign flip and a permutation between the $\mathrm{A}$ and $\mathrm{B}$ sites:
%
%
 \begin{align}\label{simpleham2}
  G h_{\mathrm{DC}} G^{-1}= 
     \begin{pmatrix}
        0 & -t_{\mathrm{B}\rightarrow\mathrm{H}} & 0\\
       -t_{\mathrm{H}\rightarrow\mathrm{B}} & 0 & -t_{\mathrm{H}\rightarrow\mathrm{A}}\\
        0 & -t_{\mathrm{A}\rightarrow\mathrm{H}} & 0
     \end{pmatrix}
\end{align}
\bibliography{bibliography}

\end{document}